# Dynamical Modeling of NGC 7252 and the Return of Tidal Material

J. E. Hibbard[1,2]

Astronomy Department, Columbia University

538 W 120 Street, New York, NY 10027

*hibbard@galileo.ifa.hawaii.edu*

J. Christopher Mihos

Board of Studies in Astronomy and Astrophysics

University of California, Santa Cruz, CA 95064

*hos@lick.ucsc.edu*

## Abstract

Motivated by recent neutral hydrogen observations with the VLA, we have undertaken an investigation into the interaction that produced the well known merger remnant NGC 7252. Through fully self-consistent N-body simulations, we are able to reproduce the kinematic character of the H I observations quite well, including the velocity reversals observed along each tidal tail. In the simulation these reversals arise from particles which have turned around in their orbit and are moving to smaller radii. The bases of the tails fall back quickly to small pericentric distances, while the more distant regions fall back more slowly to ever increasing pericentric distances. The delayed return of tidally ejected material may extend over many Gyr.

The evolution of this merger is followed numerically for 800 $h^{-1}$ Myr beyond the best fit time. We find that nearly half of the present tail material, or of order $10^9 \, h^{-2} \, M_\odot$ of neutral hydrogen and $2 \times 10^9 \, h^{-2} \, L_\odot$ of starlight, will return to within 13 $h^{-1}$ kpc of the nucleus within this time span. While the collisionless stars of the tails will continue orbiting between their inner and outer turning points, the observations show the H I gas of the tails disappearing upon its return. We discuss this result in light of the lack of central H I in the main body of this merger remnant.

*Subject headings:* galaxies:individual (NGC 7252), galaxies:interactions, galaxies:mergers, galaxies:kinematics and dynamics

---

[1] Hubble Fellow

[2] Now at Institute for Astronomy, 2680 Woodlawn Drive, Honolulu, HI 96822

# 1. Introduction

An extensive body of research has been amassed on NGC 7252, the prototypical merger remnant (Schweizer 1978, Toomre 1977). Early work showed that this remnant has a pair of counter-moving tidal tails and a single nucleus (Schweizer 1978); that the remnant light is characterized by an $r^{1/4}$ light profile and has a post-starburst nuclear population (Schweizer 1982); and that it obeys the Faber-Jackson relationship (Lake & Dressler 1986). Further studies have all reinforced the idea that this system is a late-stage merger of two gas-rich disk galaxies (Schweizer 1990; Dupraz *et al.* 1990, Wang, Schweizer, & Scoville, 1992; Fritze-von Alvensleben & Gerhard 1994; Hibbard *et al.* 1994, hereafter referred to as HGvGS). In particular, the recent VLA observations of the neutral hydrogen distribution (HGvGS) show that the tidal tails are rich in H I, while the remnant body has a surprisingly low atomic gas content. Since the optical properties of the remnant make it indistinguishable from an elliptical in the "fundamental plane", while the gas-rich tidal features require late-type spiral progenitors, this system provides the strongest observational support for the hypothesis that at least some disk galaxy mergers can produce elliptical galaxies (Toomre & Toomre 1972).

A numerical model of this system was constructed by Borne & Richstone (1991) using a "multiple 3-body" code (Borne 1984), and was re-run with a fully consistent N-body code by Mihos, Bothun, & Richstone (1993). These early numerical models featured a retrograde encounter geometry (*i.e.* the spin of the progenitors' disks was opposite to their mutual orbital motion) in order to match the anomalous velocities observed in the ionized gas along the minor axis (Schweizer 1982). Furthermore, the modeled encounter geometry required a very tightly bound orbit, such that the galaxies were in contact at the start of the simulation. The conditions which would give rise to such a tightly bound orbit were left unresolved. While these simulations produced a final remnant that was consistent with the morphology and measured velocities at the time, they are inconsistent with the new H I data (HGvGS). In particular, they give a velocity gradient along the tails of the wrong sign.

The new VLA observations map the tidal velocity field in two-dimensions, sampling the kinematics of the tails in a significantly more complete manner than the H$\alpha$ observations of Schweizer (1982). They are therefore a more reliable record of the past encounter geometry than the the inner gas kinematics. These observations reveal that the line-of-sight velocities reverse their sense of motion (relative to systemic) between the ends and the apparent bases of the tails. Since the ends of the tails are most likely kinematically expanding (Toomre & Toomre 1972), these motions are interpreted by HGvGS as an indication that the material within the base of the tails has turned around in its orbit and is now moving to smaller radii. They proposed that a prograde, initially parabolic encounter geometry may reproduce both the observed tail kinematics and the velocity reversals in NGC 7252. The present study tests this hypothesis by using the morphology and kinematics of the tails, as mapped by the neutral hydrogen, to constrain fully self-consistent, 3-dimensional N-body numerical models of the encounter. Since the tails are rich in atomic gas, their kinematics are uniquely suited for comparisons with numerical models.

The idea that tidal material remains bound to the merger remnant and returns at late times has been seen repeatedly in numerical simulations of mergers (Barnes 1988, 1992, Hernquist 1992, 1993a, Hernquist & Spergel 1992). The observations of HGvGS are the first time material has been observed with the kinematical signature of this return. Since then, the merger remnant NGC 3921 was observed to exhibit similar tail kinematics (Hibbard & van Gorkom 1995). This return is believed to be especially important for the atomic gas. In both NGC 7252 and NGC 3921 the remnant bodies remain largely H I free, and the column density of of H I is seen to drop sharply along the tails towards the remnant body, after the observed velocity reversals. In NGC 7252, the gas column density drops by a factor of 6 in a projected distance[1] of 3 $h^{-1}$ kpc at the base of the NW tail (HGvGS). At the observed line-of-sight velocity of 100 km s$^{-1}$, this corresponds to a rate of H I depletion of 1-2 $M_\odot$ yr$^{-1}$. Although the present simulation does not include hydrodynamical effects, the kinematics of this material may shed some light onto how the remnant body and surrounding shells and loops become and remain largely free of atomic gas, and why the atomic gas density falls off towards the remnant body.

We are also interested in the ultimate fate of the present tidal material. In NGC 7252, HGvGS find $2 \times 10^9 \, h^{-2} \, M_\odot$ of H I and $3 \times 10^9 \, h^{-2} \, L_\odot$ of starlight within the tails. This is a substantial amount of material, and its ultimate distribution may have important implications for the future evolution of this system. How much will fall into remnant, and on what timescales? How much is unbound? To help answer these questions, we follow the final simulation for twice as long as the best-fit time.

This paper is structured as follows: §2 describes the modeling technique; §3 compares our best fit simulation to the H I observations; §4 examines the kinematic structure of the tails and future evolution of the best fit model; §5 discusses the implications for the future evolution of NGC 7252, and finally, §6 presents our conclusions.

## 2. Numerical Technique

In order to define a unique orbital interaction solution, a large number of parameters must be defined, including (but not limited to) the mass ratio of the galaxy pair, the orbital parameters of the encounter (energy and angular momentum), the orientation of the disks to the orbital plane, and the observing time and viewing geometry (*e.g.* Mihos, Bothun, & Richstone 1993). Because NGC 7252 is in the final stages of a merger, much information concerning the structure of the progenitor galaxies has been lost, such as their rotational speeds, luminosity profiles, and the relative luminosity ratio between the pair. Rather than attempt an exhaustive and expensive search of such a vast parameter space, we make a number of simplifications and assumptions about the progenitor structure and interaction geometry to streamline the model matching process.

---

[1] All length, mass, and luminosity scales in the following are expressed in units involving the scale factor $h$ defined in terms of the Hubble constant $H_0$ via the relationship $H_0 = 100 \, h$ km s$^{-1}$ Mpc$^{-1}$

The fact that NGC 7252 shows two long, opposed tidal tails suggests that the progenitor galaxies were disks of comparable mass, as the formation of such tidal features is most easily achieved through an interaction of equal-mass disk galaxies (Toomre & Toomre 1972). Therefore, we model the initial galaxies as identical two-component systems, consisting of a luminous disk and a dark matter halo (see Hernquist 1993b). The disks feature an exponential density profile with radial scale length $h_r = 1$ and total mass $M_d = 1$. The dark halos are described as truncated isothermal spheres of mass $M_h=5.8$ and core radius $r_c=1.0$. The combined mass distribution of each disk-halo pair yields a rotation curve which shows a smooth rise between $0 < R < 2$, stays reasonably flat from $2 < R < 6$, and declines gradually at $R > 6$. The final scaling of the model units to physical quantities will be set by matching the model distances and radial velocities to the angular scales and velocities of the observations.

We make two further assumptions concerning the orbital parameters of the encounter. Unlike the tightly bound elliptical orbit in the model of Borne & Richstone (1991), we take the initial orbit to be parabolic. Galaxy encounters which are strongly hyperbolic will likely not merge due to the inefficient transfer of orbital energy to internal energy, while tightly bound elliptical orbits suffer the opposite problem – their short decay times make them less apt to survive for a Hubble time and be observed merging only now. Finally, we choose the perigalactic separation of 2.5 disk scale lengths, in order that the progenitors merge on a timescale of order 1 Gyr (Hernquist 1992, 1993a). This is the approximate age of the NGC 7252 encounter based on the distances and velocities of H II emission regions near the ends of either tidal tail (Schweizer 1982).

With these four choices made (progenitor structure, mass ratios, parabolic encounter, and perigalactic separation), our task is reduced to finding the best set of disk orientations, viewing angles, and time since orbital periapse[2] that best match the H I observations. The orientation of the disks are described by the angle pair $(i,\omega)$, where $i$ is the inclination of the disks' angular momentum vector towards ($i > 0$) or away from ($i < 0$) the companion at orbital periapse, while $\omega$ is the argument of periapse (*c.f.* Fig. 6 of Toomre & Toomre 1972).

As Borne & Richstone (1991) explicitly demonstrated, a cross-tailed morphol-ogy is obtained by many encounter geometries, and it is the kinematics which provides a strong constraint on the models. We differ from their earlier modeling effort in that we use the new H I velocity field of the outer regions ($30'' < r < 400''$) to constrain our models rather than the kinematics of the inner ionized gas disk ($r < 35''$) as measured along two slit position angles by Schweizer (1982), which is given no weight. This explains why our solution, while morphologically similar to Borne & Richstone (1991), is so kinematically different.

---

[2] We use the term "orbital periapse" to denote the point of closest approach of the progenitors on the original parabolic orbit. We will use the term "pericenter" when referring the inner radial turning point for individual simulation particles.

We believe we are justified in our approach for the following reasons: Since tails develop kinematically from the anti-tidal regions of the outer disks (Toomre & Toomre 1972), they remain beyond the most violently changing regions of the potential. Additionally, the low column densities seen in the tails (HGvGS) makes it unlikely that hydrodynamical effects are important. The tail kinematics will therefore be determined by the original trajectories of the progenitors and the global potential at large radii, and should be largely unaffected by the fine details of the final merger, making them a reliable tracer of the encounter history. The inner regions, on the other hand, are many dynamical times old, have higher local densities, and are in close proximity to the violently changing regions of the potential and any central star burst. As such, they are much more sensitive to the precise details of the final galaxy coalescence, which is only roughly modeled, and are more likely to be affected by hydrodynamical effects.

For similar reasons, we make no attempts to match the many fine structure features (shells, loops, and ripples) seen around the remnant (see figures in Schweizer, 1982, and HGvGS). The morphology of these features depends sensitively on the initial structure of the progenitors, the details of the final merger, and such things as the number of particles and smoothing length used in the simulation. However, since they contribute only a small fraction of the total light of the remnant (HGvGS), they will have little if any dynamical effect on the evolution of the tails.

Our first set of simulations designed to narrow down parameter space was done with a restricted 3-body code, similar to that used by Toomre and Toomre (1972), using massless test particles and rigid potentials for each galaxy. However, the code was written to force the potentials representing the galaxies to follow merging trajectories calculated from a full N-body parabolic encounter. While such models are unable to describe effects such as orbital decay and bar formation in the disks, they yield reasonable models for the morphology and kinematics of the tidal tails (Borne & Richstone 1991). The resulting tidal tails produced in these models were compared with the observed tail morphology and general velocity structure of NGC 7252. From this survey of over 50 different encounter geometries, it was found that a prograde encounter was required to get both the tail velocities and velocity gradients with the correct signs. The allowed inclinations were restricted to a 30° range and each of the arguments of periapse were restricted to a 50° range in order to produce tails with roughly the correct angle between them and with the proper sense of velocity.

We further narrowed phase space by using a full N-body treecode (Barnes & Hut 1986, Hernquist 1987), but with only 2048 particles per galaxy. At this stage, the simulation particles were compared directly to the H I morphology and velocity structure (as described in §3 below). The spatial and velocity scaling parameters were changed along with viewing angles to best match the H I data, as judged by eye (for details see Hibbard 1995). The primary factor in these fits was the requirement to produce the velocity reversal along the NW tail while giving tails of the proper morphology. After 16 runs we had further restricted the allowable range of disk inclinations and arguments of periapse to within 10° and 30° of the final values, respectively. We then iterated around disk geometries in this restricted phase space using a larger N-body simulation, utilizing 8192 particles per galaxy. After 5 runs, we arrived at the solution described below. This final solution was repeated with a total of 65,536 particles (16,384 per disk, 16,384 per halo) to reduce the compromising effects of $\sqrt{N}$ noise on the dynamics and particle statistics.

Our best fit model involves two equal mass disk galaxies on an initially parabolic orbit with an orbital periapse of 2.5 disk scale lengths. The disk which gives rise to the northwest tail has an orientation of $(i,\omega) = (-40,0)$ while the disk forming the east tail has an orientation of $(i,\omega) = (70,-40)$. Taking orbital periapse as the time zero-point, the simulation is started when the galaxies are 30 disk scale lengths apart, at T=-24.

While we have made no attempt to fit the observations using differing orbits or dark matter distributions, we have run a series of simulations to determine experimentally how such parameters affect the properties of the final system. In one set of calculations we kept the orbital periapse fixed while changing the compactness of the dark matter halos, while in a second we held the halo structure fixed and varied the orbital periapse. In summary, galaxies with diffuse halos merge more slowly and display very diffuse, low mass tidal tails, unlike those observed in NGC 7252, while galaxies with compact halos merge quickly and possess more massive tidal tails. Galaxies merging on a wider orbit than that used in our model experience a less violent perturbation at periapse and therefore they sport rather anemic tidal tails which are greatly dispersed by the time the disks merge. A combination of wider orbit and more compact halo may result in a merger geometry similar to that of our best fit models; however, galaxies with such compact halos will show rapidly declining rotation curves, unlike disk galaxies observed today. Given the rather weak constraints on the orbit and halo structure from the extant data, we feel our model represents a good compromise between the need to survey a wide range of parameter space and the desire to have astrophysically "reasonable" initial conditions.

## 3. Simulation Fit to the H I Data.

The VLA[3] observations by HGvGS produce a data cube of H I intensities gridded in right ascension and declination ($\alpha$, $\delta$) for each velocity "channel". We use the BnC-array subset of these data, which has the best spatial and velocity resolution (beam full width at half maximum of $30''\times 30''$, velocity channel spacing of 10.5 km s$^{-1}$; see HGvGS for details). The kinematics of the observed H I emission is presented in figure 1, reproduced from Fig. 4 of HGvGS. Fig. 1a presents the intensity-weighted mean H I velocity field of the entire galaxy, superimposed upon a negative greyscale image of the H I column density distribution. The three main components of H I are labeled: the northwest (NW) tail, the eastern (E) tail, and the western (W) loop complex. A special note is made of the gas at the apparent base of the NW tail which is red-shifted with respect to the systemic velocity of 4740 km s$^{-1}$. Fig. 1b shows a position-velocity diagram for this system directly below Fig. 1a and on the same spatial scale as it. This map has been constructed by summing the gas emission along the declination axis, and plotting the integrated intensity at each velocity ($y$-axis) as a function of the differential right ascension to the remnant center ($x$-axis).

This mapping provides a clear picture of the smoothness and completeness of the velocity field, and shows that all of the detected H I is associated with one of the three main kinematic features. The E tail is red-shifted relative to systemic, while most of the NW tail is blue-shifted. The absolute value of the relative velocity rises as one traces the H I outward from the center along each tail, reaches a maximum, and then falls back again. Near the apparent base of the NW tail, the velocities are higher than systemic by up to 130 km s$^{-1}$ (*i.e.* red-shifted relative to center), whereas the rest of this tail has velocities lower than systemic. To the west (W) of the remnant center is gas associated with a bright star forming loop. This gas is blue-shifted by more than 200 km s$^{-1}$, and its morphology and kinematics suggest that it crosses in front of the remnant in the north, passes behind it to the south, and that it may connect back to the eastern tail.

To represent the gas kinematics, we use a subset of these data taken from the "clean components". The clean components are the locations ($\alpha, \delta$) and intensities of a set of delta functions at each velocity channel (centered at $V_{los}$), which, when convolved with the array spatial sampling function, reproduce the observations to within the noise (*c.f.* Clark 1980). Because extended structures need to be represented by many such delta functions, we only consider clean components taken from within tight boxes set around each emission feature in the individual channel maps. We further filter these components by imposing a flux cutoff of 0.9 mJy beam$^{-1}$ (equal to 3 times the r.m.s noise of the final maps), and taking no more than 6 components per box. This procedure results in 135 sets of ($\alpha, \delta, V_{los}$) data points spread over 32 velocity channels. These points are only used as markers of the gas kinematics, and do not fully map out the faint H I distribution. They do however fully represent the kinematics of the structures seen in Fig. 1. We also note that, aside from the flux cutoff used in their selection, the intensities of the clean components are in no way used to constrain the simulations.

---

[3] The VLA of the National Radio Astronomy Observatory is operated by Associated Universities, Inc., under cooperative agreement with the National Science Foundation.

**FIGURE 1:** The H I kinematics of NGC 7252, from the data of HGvGS. Both panels are on the same angular scale and the major kinematic features are labeled. **(a)** The iso-velocity contours overlaid on a negative greyscale representation of the atomic gas column densities in the range $3$–$30 \times 10^{19}$ cm$^{-2}$. The crosses mark the location of the remnant center and three H II regions, one in either tail and one in the W loop, from Schweizer (1982). The H I synthesized beam is indicated in the lower left. **(b)** Position-velocity profile, obtained by integrating the H I along the declination axis. The projection of the angular distance from the center along the right ascension is plotted on the $x$-axis and the velocity relative to systemic (4740 km s$^{-1}$) is plotted along the $y$-axis. The lines through the origin mark the spatial and kinematic center of the remnant as defined by the $^{12}$CO($1 \to 0$) measurements of Wang *et al.* 1992. Diamonds mark the location of the H II regions, with the width and height indicating the uncertainties in position and velocity, respectively. These plots illustrate the continuity of the H I features in space and velocity and show that all the detected atomic gas is associated with one of three main structures: the two tidal tails and the W loop complex. Note that the NW tail goes from being blue-shifted in its outer parts to having red-shifted velocities at its apparent base.

The simulations produce the 6-dimensional phase-space coordinates ($X, Y, Z, V_x, V_y, V_z$) for each particle at regular time-steps [4]. These data are projected directly onto the filtered clean components in three orthogonal planes: the sky plane ($\alpha, \delta$), and two position-velocity planes: ($\alpha, V_{los}$) and ($\delta, V_{los}$). The simulation points are rotated and stretched to best match the H I morphology and velocity structure in these three planes.

The projection of the best fit model run onto the clean components is shown in figure 2. The sky plane is shown in the top left of this figure, the ($\alpha, V_{los}$) plane is drawn below this, the ($\delta, V_{los}$) is drawn in the top right, and the unobserved "Top" view ($\alpha, z$) in drawn in the lower right. The clean components are indicated by open circles in the first three planes, while simulation particles are drawn as small black squares or points in all four planes. The left-hand side of this figure, minus the simulations particles, is analogous to both panels of Fig. 1. The clean components appear less continuous in Fig. 2 than the H I emission in Fig. 1 due to the filtering described above. For clarity, a subset of the simulation particles are drawn in this figure. This subset includes all tail particles ($r > 100''$), every 10th particle with a radius between $r = 25''$ and $r = 90''$, and every 25th particle within $r = 25''$. There is a gap between $90''$–$100''$ from which we select no particles. This was done for display purposes only, to separate the base of the tails from particles associated with the remnant body.

The best match to the observation data occurs at time T=72. In this fit, the system is observed $\sim 35°$ above the orbital plane and $\sim 35°$ in front of the line connecting the galaxies at orbital periapse. The best fit velocity and spatial scaling factors are 180 km s$^{-1}$ and 1.5 $h^{-1}$ kpc, respectively. This sets the time unit to 8.0 $h^{-1}$ Myr, and our best match time corresponds to $\sim 580$ $h^{-1}$ Myr after periapse of the initial orbit. These scalings result in progenitors with radial scale lengths of 1.5 $h^{-1}$ kpc and masses of $7.5 \times 10^{10} h^{-2} M_\odot$, and a mass-to-blue light ratio of the remnant at the current time of 5 $M_\odot L_\odot^{-1}$. The total mass and disk scale length imply that the progenitors were small spirals.

We find 6% of the disk particles in the tails at the best match time, with 4% in the NW tail and 2% in the E. This is very close to the observed value of 7% of the blue light found in the tails by HGvGS, with 5% in the NW tail and 2% in the E. These numbers should not be directly related to the fractional H I content of the tails, since the H I has apparently been removed from the inner regions of the remnant during the merger (HGvGS; see also §5.1 below). In the model, 8% of the tail particles or 0.4% of the total luminous matter has positive energy. Therefore,

---

[4] In the following, we will label plots measured in the orbital plane of the two galaxies with the coordinates ($X$, $Y$, $Z$), and those in the sky plane with ($\alpha$, $\delta$, $z$).

**FIGURE 2:**  Final fit from the simulation projected onto the filtered VLA clean components (see text). The clean components are indicated by large open circles, and are used to trace the morphological and kinematic structure of the gas. The simulation tail particles are drawn as small black dots, and the other simulation particles are drawn as points. The left hand panels, minus the simulation particles, are a subset of the data presented in Fig. 1. The model succeeds in reproducing both the observed spatial morphology and the general velocity structure of the tails, including the sign, magnitude, and gradient of the velocities along most of length of the tails.

99.6% of the luminous matter in the present model remains bound to the remnant. We find that 6% of the dark matter from each system is unbound, similar to what is found in other studies (Barnes 1992, Hernquist 1992).

The model succeeds in reproducing both the observed spatial morphology and the general velocity structure of the tails, including the sign, magnitude, and gradient of the velocities along most of the length of the tails. Unlike earlier models of NGC 7252 which were driven towards tightly bound, retrograde orbits (Borne & Richstone 1991; Mihos, Bothun, & Richstone 1993), our model describes a more astrophysically reasonable merger scenario which better matches the observed system morphology and tail kinematics. This was made possible by the more complete mapping of the tail kinematics afforded by the H I observations.

**FIGURE 3:** (a) The CTIO 4m *B*-band image of NGC 7252 from HGvGS. (b) The best fit projection of the simulation data at T=72 or 580 $h^{-1}$ Myr since orbital periapse. A clump that develops in the NW tail is clearly seen. Since this clump arises from statistical noise, its location in the tail is purely random.

Still, some discrepancies between the model and observations remain. Most noticeably, the velocities at the base of the NW tail are only one-half of the observed values, and the velocity structure of the E tail differs from the model in detail. Additionally the E tail is noticeably longer in the simulation, and bends slightly south as it approaches the body, while the observations suggest that it ought to continue straight behind the remnant body and bend north. However, our experience with the model matching leads us to believe that fine-tuning the initial conditions could alleviate many of these deficiencies, without significantly changing the global dynamics. Given the present uncertainties in the dark matter distribution in galaxies, such fine tuning seems unwarranted. We therefore find the present solution a plausible representation of the general dynamics of this system.

We close this section by presenting a view of the final simulation data along side the $B$-band CTIO[5] 4m CCD image from HGvGS in figure 3. This figure shows the fine match to the outer tidal morphology, but a poor reproduction of the many loops, shells, and ripples. This is not seen as a failure for the reasons mentioned in §2: the inner regions are much more dependent on the precise details of the final coalescence of the progenitors and are more susceptible to the detailed mechanics of the model.

## 4. Properties of the Simulation Model

In this section we examine the kinematic structure of the tidal tails in some detail in order to better understand their orbital evolution. §4.1 presents the time evolution of the encounter from 200 $h^{-1}$ Myr before orbital periapse to 1.4 $h^{-1}$ Gyr after, with special attention to the formation and development of the tidal tails during this period. The behavior of various kinematic parameters along the tidal tails are also examined in an attempt to understand the present kinematic structure of the tails. We find the energy and angular momentum have a monotonic relationship along the tails, leading to a relationship between the radial period and the pericentric distance. In §4.2 we use this relationship to predict the fate of the present tail material over the next few Gyrs, and describe the future evolution of the merger remnant.

### 4.1 Time Evolution and Orbital Structure of the Tails.

Figure 4 plots the time evolution of the luminous (disk) particles in the plane of the sky $(\alpha, \delta)$, from T=-16 (130 $h^{-1}$ Myr before periapse) to the best match time of T=72 (580 $h^{-1}$ Myr after periapse). The time increment between successive frames is 8 time units (64 $h^{-1}$ Myr). The progenitor of the NW tail is on the left at T=-16, while the progenitor of the E tail is on the right. The tails form soon after orbital periapse, after which the galaxies abandon their original parabolic trajectories.

---

[5] Cerro Tololo Inter-American Observatory of the National Optical Astronomy Observatories is operated by the Association of Universities for Research in Astronomy, Inc., under cooperative agreement with the National Science Foundation.

The progenitors fall together a second time at about T=36 (290 $h^{-1}$ Myr), totally losing their separate identities apart from the tails they launched.

At late times, several clumps can be seen in the expanding tidal debris, with the largest bound clump occurring in the NW tail (see the last frame of Fig. 4). The formation of these types of substructures has been covered elsewhere (Barnes & Hernquist 1992) and will not be pursued further here. We note that the clumps likely arise due to self-gravity acting upon $\sqrt{N}$ particle noise, and their location within the tails is statistical rather than physical.

To understand the kinematic development of the tails, we examine individual orbits, energy, angular momentum, and radial velocity along the tail, all at T=72. The orbital trajectories of eight particles at different distances along the NW tail are shown in figure 5a as viewed from above the orbital plane $(X,Y)$. In the accompanying panels we plot the (b) energy, (c) radial velocity with respect to the remnant center ($\dot{r}$), and (d) angular momentum, each as a function of radius. Tail particles are indicated by solid dots, while non-tail particles are drawn as points.

Examining Fig. 5a, we see that the trajectories of the outermost tail particles lie perpendicular to the tail and point outward ($\dot{r}>0$). Further in, however, the particles have reached the apex of their orbits, after which they stream along the tails, with their orbits parallel to its length and pointing inward ($\dot{r}<0$). We find that the radial velocities switch sign at a radius of about 200″ (∼45 $h^{-1}$ kpc), indicated by the dashed lines in Fig. 5c. It is the existence of this "turn-around" radius (Barnes 1988) which leads to the kinematic structure observed in NGC 7252 and shown in Figs. 1&2, whereby the line-of-sight velocities with respect to systemic change sign as one moves along a tidal feature.

The most important characteristic of the energy and angular momentum distributions shown in Fig. 5 is their monotonic behavior with distance along the tail. The bases of the tails are more tightly bound and of lower angular momentum, while the ends are of both higher energy and angular momentum. The monotonic behavior is a direct result of the kinematic development of the tails, whereby the particles sort themselves in radius according to their energy and angular momentum. These plots become double valued for features which have fallen through the potential and are moving to larger radii on the opposite side creating shell features (the lower energy and angular momentum material seen in the lower left of Fig. 5b&c out to about 150″), but are single valued along the tails. The increased scatter seen especially in Fig. 5c at r=230″ are particles within the large self-gravitating clump which develops within the NW tail, which experience small torques from the local density enhancement.

We find that once the tails are launched *they are not simply in free expansion*. On the contrary, since the bases of the tails are more tightly bound than the ends, they turn around rather quickly in their orbits and fall back[6] towards the remnant. The

---

[6] We use the term "fall back" in the present context to indicate that the tidal material moves to smaller radii. The simulation particles will start moving outward again once they reach their inner radial

**FIGURE 4:** The time evolution of the two disks as seen in the sky $(\alpha, \delta)$. The time in model units as measured from orbital periapse is given in the upper right hand corner, and the increment between frames is 8 time units (64 $h^{-1}$ Myr). The best fit to the data occurs at T=72. The progenitor of the NW tail appears to the left of this figure at T=-16, while the progenitor of the E tail on the right.

---

turning points.

**FIGURE 5:** The orbital structure of the tidal tails at the present epoch. Tail particles are drawn as solid dots while non-tail particles are drawn as points. **(a)** The projection of the NW tail onto the orbital plane $(X_o, Y_o)$ at T=72. Eight particles are numbered by increasing distance along the tail, with their past and future orbital paths shown. The remaining panels show behavior of the various orbital quantities with distance along the tail at T=72. **(b)** Energy as a function of radius. **(c)** Radial velocity as a function of radius. The vertical dashed line shows the intersection with the zero crossing, and therefore defines the "turn-around" radius. moving inward. **(d)** Angular momentum with radius. These plots show that radial velocity, energy, and angular momentum are single valued and monotonically increasing with distance along the tail. As a result, period and pericentric distance will also be single valued and monotonic with radius.

more distant material is less bound and thus falls back more slowly, if at all. The rate of return falls off roughly as $t^{-\frac{5}{3}}$, as expected for a flat distribution of binding energies (Toomre 1977, Barnes 1992). This behavior implies a much higher rate of return initially, and a continued, lesser rate of return far into the future.

We illustrated the above behavior by color coding the material in the NW disk according to radius soon after periapse. This is done in figure 6, where we repeat the first nine frames of Fig. 4 as viewed now in the orbital plane $(X, Y)$. Magenta is used for the innermost, most tightly bound disk (*i.e.*, non-tail) material, blue is the material next farthest out, followed by green and finally red. Luminous tails containing one-quarter of the total disk material are raised initially (T=16 ~130 $h^{-1}$ Myr, blue, green and red material). However, by T=32 (260 $h^{-1}$ Myr) nearly half of this material has fallen back in (blue material), mixing extensively with the innermost disk material. Over one-half of the remaining material will fall back in by T=72 (580 $h^{-1}$ Myr, green). As a result, the current NW tail contains only one-fourth of the material raised initially, and tidal material continues to fall back today.

The return of tidal material is a direct result of most of the tail material remaining bound (Barnes 1988, 1992, Hernquist 1992, 1993a, Hernquist & Spergel 1992). The most tightly bound disk material experiences a rapidly changing potential of the merging galaxies, and scatters onto a wide range of orbits (Barnes 1994). This violent relaxation is thought to give rise naturally to a $r^{1/4}$ light profile (Lynden-Bell 1967, van Albada 1982), and sets up a much more regular mass distribution. Particles which fall back after the potential has relaxed somewhat (*i.e.* for which the fall-back time is roughly equal to the dynamical time) will turn around and head back towards their outer radial turning point. After this time, the material which continues to rain down upon the remnant can wrap coherently, producing loops, shells, ripples, and other fine structure features (Hernquist & Spergel 1992). In this simulation, the first shells form around T=48 (390 $h^{-1}$ Myr) from material that passed nearly radially through the center of the potential at T=40 (green particles in Fig. 6).

The subsequent evolution of the tails is explored by running the simulation beyond the best match time, to T=172 or 1.4 $h^{-1}$ Gyr since first periapse. Very little happens in the remnant body and so for clarity we show only the evolution of the particles with r>100″ at T=72. The evolution from T=72 to T=168 (0.8 to 1.3 $h^{-1}$ Gyr) is shown in figure 7 looking down onto the orbital plane. A circle of radius 57″, or four times the measured effective radius of NGC 7252 ($R_e$=14.3″, HGvGS), is drawn in the first and last panels of Fig. 7. This figure illustrates quite well the fate of the even later returning tail material. Soon after these streams of material fall back into the remnant, they turn around and move back out, creating loop features. Such loop structures are seen in NGC 7252, and the morphology of the loops in the present simulation lends support to the suggestion by Schweizer (1982) that the eastern tail passes behind the face of the remnant, connects to the western loop complex in the south, and crosses in front of the remnant to the north. The simulation also supports Schweizer's (1982) contention that the NW arm crosses the face of the remnant and connects to a prominent loop in the south-southeast quadrant of the remnant.

**FIGURE 7:** The future orbital evolution of the tail particles alone, shown looking down onto the orbital plane $(X_o, Y_o)$. The time in model units as measured from orbital periapse is given in the upper right hand corner, and the increment between frames is 12 time units (96 $h^{-1}$ Myr). This figure illustrate quite well the late return of cold streams of tail material. A circle indicating $4 \times R_e$ is drawn in the first and last frames. The returning material forms loops similar to those seen in NGC 7252.

The tail doubles its overall length by the end of the run, reaching over 200 $h^{-1}$ kpc, and increases its total projected area by a factor of 6. However, while some regions experience similar increases in radius and cross-sectional area (notably at the freely expanding ends of the tails), many regions experience much smaller increases. This is due to the deceleration experienced by the bound particles. As an illustration we examine the evolution of points 5&6 in Fig. 5a. We find that as time increases by 140% (from T=72 to T=172), these points increase their projected separation from the remnant center by only 20%, and the surface area of the tail between these points grows by only a factor of two rather than the factor of six expected from free expansion. In addition, self-gravity of the tidal material causes some regions to stick together, at the expense of adjoining regions. As a result of these effects, the column density of tidal material will not drop evenly across the tails, and we expect them to have a clumpy appearance in the future.

### 4.2 Future Evolution of NGC 7252

To understand the evolution and fallback of tidal material at late times ($> 1.5$ Gyr), we wish to determine the orbital periods and pericentric distances of the tail particles. This is possible at late times and at large radii because once the remnant becomes relaxed, the individual energies and angular momenta of the tail particles are approximately conserved. We find that the energies are constant to within $\pm 7\%$, and that the angular momenta vary by less than 15% from T=72 to the end of the run at T=172. As a result, we can use these quantities to calculate the period and pericentric distance of the tail particles, in order to predict roughly when and how closely they will approach the remnant.

Rather than the costly approach of evolving the simulation indefinitely to determine these quantities, we turn to an analytic extension of our numerical model. We approximate the mass distribution of the merger remnant by a spherical analytic form, allowing the orbital elements for the tidal material to be derived simply as a function of energy and angular momentum. We model the mass distribution in the remnant using a two-component Hernquist (1990) potential-density pair:

$$\Phi(r) = -\frac{M_{lum}}{r + a_{lum}} - \frac{M_{dark}}{r + a_{dark}},$$

where the subscripts *lum* and *dark* refer to the luminous material and dark matter halos, respectively. $M_{lum}$ and $M_{dark}$ are simply the mass in each component ($2.1 \times 10^{10} h^{-2} M_\odot$ and $1.3 \times 10^{11} h^{-2} M_\odot$, respectively), while the radial scale length $a$ is calculated using the relationship $a = (1 + \sqrt{2})r_{\frac{1}{2}}$ for the Hernquist potential-density pair. We calculate $a$ for each component at T=124, and derive $a_{lum} = 0.9$ and $a_{dark} = 4.5$ (1.3 $h^{-1}$ kpc and 6.5 $h^{-1}$ kpc, respectively). These values change by $\lesssim 10\%$ from T=72 to T=172 (the end of the $N$-body simulation).

The adopted potential is spherical, and any non-spherical component to the remnant may complicate the orbital evolution of the tails over that described by our analytic model. However, the dark matter halo is nearly spherical (half mass axis ratios 1:0.95:0.8), and comprises 85% of the remnant mass, suggesting that any discrepancies in the orbital evolution will occur mainly at small radius where the luminous component (half mass axis ratios 1:0.7:0.6) dominates the mass distribution. To test the suitability of the analytic fit, we have evolved the tail particles from T=72 to T=172 using only the analytic potential, and find a good match between the evolution of these particles compared to the full $N$-body run. The analytic model provides reasonable estimates, therefore, for quantities such as the orbital period and pericentric radius ($R_{peri}$) for the tail material.

We plot these quantities against each other in figure 8a. The pericentric distance is quantified in terms of the number of effective radii ($R_e$) by dashed horizontal lines. Arrows are drawn at the top of this plot to indicate the period for particles which will reach pericenter by the end of the simulation ($t_{end}$=1.4 $h^{-1}$ Gyr), and the period of particles which will reach their pericenter at 3 $h^{-1}$ Gyr since orbital periapse. The particles are color coded according to their return times and radii. Magenta denotes particles with a pericentric distance of less than $4R_e$ and a period shorter than 1.4 $h^{-1}$ Gyr. In other words, all of these particles should penetrate to within $4R_e \approx 13$ $h^{-1}$ kpc by the last timestep of the simulation. Blue and green particles are those which are predicted to return to within $4R_e$ and $5R_e$ respectively between 1.4 and 3 $h^{-1}$ Gyr. Red particles are expected to approach within $5R_e$ sometime between 3 $h^{-1}$ Gyr and a 10 $h^{-1}$ Gyr of orbital periapse. Finally, the black particles are expected to remain at radii greater than $5R_e$ for at least a Hubble time. We find that the black particles account for 20% of the present tail particles. Therefore, while only 8% of the current tail particles are unbound (§3), 20% are effectively lost from the remnant, in that they will not fall back to within $5R_e$ in a Hubble time.

Due to the monotonic behavior of period and angular momentum with radius, the low angular momentum, short period inner regions of the tails will have a correspondingly small $R_{peri}$ (i.e., the bases of the tails occupy the lower left of the Period–$R_{peri}$ plane), and the high angular momentum, long period regions will have a correspondingly large $R_{peri}$. This is shown in Fig. 8b, which reproduces Fig. 3b using the above color coding. The predictive accuracy of this plot is checked by comparing the number of magenta particles identified at T=72 with the actual number which are found to fall to within $4R_e$ by the end of the simulation run. Fig. 8a predicts that 45% of the tail material should return to within $4R_e$ by the end of the run, while the actual N-body integration finds that 42% actually returns.

We use the information presented in Fig. 8 to estimate the quantities of gas and starlight which are expected to fall within the remnant body of NGC 7252. We do not simply count up the number of particles of each color, as this number will depend on the actual distribution of matter in the progenitor disks (i.e., the disk scale length). This distribution is poorly constrained, and may be inappropriate for the neutral hydrogen. Instead, we use the kinematic information contained in the Period-$R_{peri}$ plot to construct a "mask" to use on the actual observational data. We then calculate the mass of H I and luminosity of star light that is contained within the different regions of the mask. For example, we use the extent of the magenta material in Fig. 8a to calculate the amounts of atomic gas and starlight predicted to fall to within $4R_e$ in the next 800 $h^{-1}$ Myr.

The results of this exercise are listed in Table 1. The first column describes the characteristics of the delineated region, in terms of its maximum pericentric distance and the periods involved. The second column lists the time as measured from orbital periapse. The third column list the time as measured from the present epoch ($t-t_{now}$). The fourth column lists the quantity of blue light contained within the delineated region. Column 5 lists the quantity of H I contained within the region, and finally column 6 lists the corresponding color used in Fig. 8. The first line of this table list the total quantities of light and atomic gas presently within the tails. In summary, we expect one-third of the H I and two-thirds of the blue light currently within the tails to fall back into the remnant over the next 800 $h^{-1}$ Myr. The stars should fall right back out, but at least at the present epoch the atomic gas does not make it this far. $4 \times 10^8 \, h^{-2} \, M_\odot$ of atomic gas and $2 \times 10^8 \, h^{-2} \, L_\odot$ of blue light are expected to remain far from the remnant ($>5R_e$) for over a Hubble time.

TABLE 1. Return of Tidal Material.

| (1) Amount of Material that ... | (2) time ($h^{-1}$ Gyr) | (3) $t - t_{now}$ ($h^{-1}$ Gyr) | (4) $L_B$ ($h^{-2} L_\odot$) | (5) $M_{\rm HI}$ ($h^{-2} M_\odot$) | (6) Color in Fig. 8 |
|---|---|---|---|---|---|
| is currently within the tails | 0.6 | 0.0 | $2.6 \times 10^9$ | $2.0 \times 10^9$ | magenta+blue green+red+black |
| falls to within $4R_e$ by the end of run | 1.4 | 0.8 | $1.7 \times 10^9$ | $0.7 \times 10^9$ | magenta |
| falls to within $4R_e$ by 3 Gyr | 3.0 | 2.4 | $2.1 \times 10^9$ | $1.0 \times 10^9$ | magenta+blue |
| falls to within $5R_e$ by 3 Gyr | 3.0 | 2.4 | $2.3 \times 10^9$ | $1.4 \times 10^9$ | magenta+blue +green |
| falls to within $5R_e$ by a Hubble time | 10.0 | 9.4 | $2.4 \times 10^9$ | $1.6 \times 10^9$ | magenta+blue +green+red |
| remains beyond $5R_e$ for over a Hubble time | 10.0 | 9.4 | $0.2 \times 10^9$ | $0.4 \times 10^9$ | black |

## 5. Discussion

HGvGS used their multiwavelength observations of NGC 7252 to show that the atomic gas lies within the tidal tails and W loop (Fig. 2), and are not projected onto these features. They argued that if the ends of the tails are predominantly expanding, then the observed velocity reversals along the NW tail are strong evidence that it remains mostly bound to the remnant, and that the base material is falling back towards the remnant body. The simulation we have conducted here confirms the plausibility of this explanation. It further illustrates that the majority of the tail material will fall back on short time scales to small radii and mix violently (Barnes 1994) while more distant material will fall back on longer timescales to larger radii, some forming loops and shells and other fine structures (see also Hernquist & Spergel 1992). The most distant material has the highest energy and angular momentum and will always remain far from the remnant.

The return of material at the present epoch has far reaching implications, which we discuss in this section. The most immediate interest to us is whether these considerations provide any explanation for the atomic gas-poor nature of the NGC 7252 remnant. We will address this question in §5.1. However, the above results are of interest beyond the specific case of NGC 7252. The return of tidal material is a generic result of merging simulations, and is therefore an expected consequence of mergers in general. In §5.2, we discuss the implications of this return.

### 5.1 Lack of Neutral Hydrogen in NGC 7252

The H I deficiency within the remnant body of NGC 7252 manifest itself in two important ways (§1 and HGvGS): (1) the main body and surrounding shells as a whole are H I poor ($M_{\rm HI}/L_B$<0.01 $M_\odot L_\odot^{-1}$); and (2) atomic gas falling towards the remnant from the tail is "disappearing" today at a rate of 1–2 $M_\odot$ yr$^{-1}$.

A lower limit to the amount of H I converted to other forms in the past is estimated by comparing the regions of the original disks which are tidally ejected with the regions which wind up in the remnant body. This is done in figure 9, where top panels present the face-on view of each disk at the start of the simulation, while the lower panels give the distribution of particles with angle. The disks giving rise to the NW and E tails are presented in the left and right columns, respectively. Particles in the present day tails are drawn with filled circles, while non-tail particles are indicated by points. The spiral pattern seen in Fig. 9 "unwinds" with the disk rotation so that the tail particles appear on the far side of the potential at pericenter (see the first three frames of Fig. 6; see also Fig. 15 of Toomre & Toomre 1972). We find that the present tails are drawn from radii $r > r_{in}$, where $r_{in} \approx 3$ $h^{-1}$ kpc for the NW tail $r_{in} \approx 4$ $h^{-1}$ kpc for the E tail (indicated by by dashed lines in Fig. 9).

This figure illustrates that *only a fraction of the disk H I can be ejected into the tails*. The remaining material ends up distributed evenly throughout the the remnant body. This material is scattered out of its originally planar orbits due to the action of violent relaxation (Lynden-Bell 1967). Since the material which was drawn into the tails from radii $r > r_{in}$ is gas rich, gas rich material must also have been sent into the remnant body. Due to the symmetric nature of the tidal force, at most 50% of the original disk material can be ejected into a tail during a merger. Therefore, given the present paucity of inner atomic gas, we conclude that at least twice as much H I has *already* been converted into other forms within the remnant body as presently resides in the tidal tails.

To get a better estimate of the amount of H I which was sent into the remnant body, we calculate the fractional disk area beyond $r_{in}$ occupied by tail particles. This calculation is rather insensitive to the choice of inner and outer radii, since the tails are drawn from "wedges" of nearly constant opening angle, $\theta$, in the original disks (lower panels of Fig. 9), and the fractional area will therefore be constant at $\frac{(r_{out}^2 - r_{in}^2) \times \theta}{(r_{out}^2 - r_{in}^2) \times 2\pi} = \frac{\theta}{2\pi}$. We find that the NW tail occupies 50% of the original disk area at $r > r_{in}$, while the E tail occupies 34%. Extrapolating from the current H I content of the tails, we conclude that *at least* $3 \times 10^9 \, h^{-2} \, M_\odot$ of atomic gas was sent inward. By extrapolating from the measured tail gas content, we avoid uncertainties in the actual radial distribution of the H I. All that we require is that this distribution be azimuthally well behaved.

**FIGURE 9:** The present day tails are indicated by filled circles according to their position in the progenitor disks at the start of the simulation (T=-24). The NW and E tail progenitor disks are presented in the left and right hand columns, respectively. The upper panels present the face on disks as viewed looking down upon their spin planes. Both systems rotate clockwise in this projection, and the spiral pattern traced out by the tail particles unwinds so that they are on the far side of the disk at orbital periapse (see Fig. 6). The lower panels present the radial distribution of particles with angle, where the angle is measured from the line of approach connecting the two systems. Dashed lines indicated the inner radial boundary for the material drawn into the present day tails. Non-tail particles are indicated by points. This figure illustrates that a large fraction of the outer disks of both systems do *not* get drawn in tails. We therefore expect at least as much H I to have been sent into the remnant bodies as now resides within the tails.

These numbers are rather strict lower limits, since it is unlikely that neither progenitor had any H I within $r_{in}$. Dividing the total $B$ luminosity between the two progenitors, we derive original H I mass-to-blue light ratios of $M_{\rm HI}/L_B \gtrsim 0.1\ M_\odot L_\odot^{-1}$, which is on the low end for the late type spirals inferred by HGvGS (Giovanelli & Haynes 1988, Young & Knezek 1989). For late type progenitors a value of $M_{\rm HI}/L_B \sim 0.3$ is more appropriate, suggesting that as much as $9 \times 10^9\ h^{-2}\ M_\odot$ of atomic gas may have been sent inward.

Since no atomic gas is found within the inner regions, HGvGS conclude that it has been converted into other forms within the remnant body over the course of the merger. The same process which redistributes the nearside disk particles, *i.e.* violent relaxation, may provide a natural explanation for the conversion of the accompanying H I into other forms. Hydrodynamical simulations show that in such a violently changing potential, the gas is unable to avoid collisions, and it dissipates energy and moves inward (Negroponte & White 1983, Barnes & Hernquist 1991, Mihos & Hernquist 1994). In these collisions, low density gas may shock-heat to temperatures where the cooling times are long, while the higher density clouds may compress into molecular form (and from there into stars). In NGC 7252 there is evidence for an extended hot gas, a post-starburst stellar population, and concentrated cold molecular gas (Dupraz *et al.* 1990, Wang *et al.* 1992, HGvGS).

We turn our attention now to the continued conversion of the retuning H I into other forms. We note in Fig. 7 that the return of the tidal material is accompanied by a noticeable orbital crowding. This orbital crowding may lead to dissipation in the gaseous material, causing it to separate from the stellar component and move inward. Our simulation would need to be repeated including hydrodynamical effects in order to address the plausibility of this scenario.

### 5.2 Return of Tidal Material

The power-law nature of the rate of return of the tidal material means that the after-effects of merging will span many Gyrs. As Figs. 4&6 illustrate, short, luminous tails are initially raised, but yield to long, less luminous features. Although these long lived tidal features contain only a small fraction of the total luminous material, they may prove to be important observationally. This is because the tails tend to be rich in atomic gas, allowing both their distribution and kinematics to be mapped over large areas (van der Hulst 1979, Yun 1992, Hibbard 1995, Hibbard & van Gorkom 1995).

The return of tidal material may provide insight into a host of observed extragalactic peculiarities. It may play a role in the formation of polar ring galaxies (Schweizer, Whitmore, & Rubin 1983; van Gorkom, Schechter, & Kristian 1987), ellipticals and S0's with extended and/or kinematically decoupled outer gaseous components (van Gorkom *et al.* 1986; Lake, Schommer, & van Gorkom 1987; Schweizer, van Gorkom, & Seitzer 1989; van Driel & van Woerden 1991), multi-epoch bulge populations (Schweizer 1990 and references therein), and high velocity clouds (van der Hulst & Sancisi, 1988). In addition, recent VLA observations of shell galaxies reveal impressive quantities of atomic gas primarily in rotation at large radii (Schiminovich *et al.*, 1994, 1995). These systems are thought to be the long term evolutionary result of such mergers (Schweizer & Seitzer 1992), and the outer H I may offer strong support for such an origin. However, to fully understand the role of mergers versus, *e.g.* multiple accretion events, more sophisticated hydrodynamical modeling is needed.

The orbital periods and pericentric distances are also of interest because the tails are often found to contain such interesting features as possible proto-dwarf irregulars (Schweizer 1978, Hibbard & van Gorkom 1993, HGvGS, Duc & Mirabel 1994). The orbital elements are necessary to determine how transient these features will be. For example, the location of the clump of gas and stars found within the NW tail of NGC 7252 has been indicated by a large black dot in Fig. 8. The period-Rperi plane predicts that this material will fall to within 14 $h^{-1}$ kpc within the next 3 Gyr. If this feature is dynamically bound, it has enough mass to avoid tidal stripping at this radius (HGvGS) and should be a long-lived companion of the remnant, orbiting between 14 and 120 $h^{-1}$ kpc with a period of about 4 $h^{-1}$ Gyr.

## 6. Conclusions

- We have constructed a numerical model for the NGC 7252 encounter which succeeds in reproducing both the observed spatial morphology and the general velocity structure of the tails, including the sign, magnitude, and gradient of the velocities along most of their lengths. The fit gives a more astrophysically satisfying orbital geometry than earlier models. Such a fit was made possible by the complete velocity mapping of outer regions afforded by the H I emission.

- Most of the tail material remains bound and is not in pure expansion. The observed velocity reversals along the tidal tails arise from material that has reached the turn-around point in its orbit and is moving to smaller radii, as suggested by HGvGS.

- The tidal material returns at an ever-decreasing rate, extending the effects of merging over many Gyr. Initially, one-quarter of the total disk material is lifted into a tail. By the present time most of this material has already fallen back and the tails contain only 5% of the original disk material. Since energy and angular momentum increase monotonically with radius along the tails, the early returning material falls to small radii and later returning material falls to larger and larger radii. These results are verified by a following the N-body simulation for 1.4 $h^{-1}$ Gyr.

- At late times and at large radii, the potential is well behaved. As a result, we are able to calculate periods and pericentric distances for material within

  the tail with some degree of confidence. By the end of the simulation half of this material returns to within $4R_e$. Within $3\ h^{-1}$ Gyr half of the remaining material is expected to return to within $5R_e$. $4\times10^8\,h^{-2}\,M_\odot$ of atomic gas and $2\times10^8\,h^{-2}\,L_\odot$ of blue light are expected to remain beyond $5R_e$ for over a Hubble time.

- The present simulation provides some insight into the low atomic gas content of merger remnants. In particular, we show that only a fraction of the disk H I can be drawn into tidal tails, and conclude that in NGC 7252 at least $3\times10^9\,h^{-2}\,M_\odot$ of atomic gas was sent inward during the merger. We suggest that violent relaxation will lead to collisions, shocks, and dissipation in the gaseous component, possibly providing a natural explanation for the conversion of this H I into other forms within the remnant body.

- The simulation suggest that the observed decrease in atomic gas along the tails towards the remnant body is associated with the first return of tidal material. While the collisionless material will move right back out in the potential after its first return, this may not be the case for the dissipational gas. Fully hydrodynamical simulations are needed to test this hypothesis.

  The remaining question is: How robust are our predictions, especially in light of the uncertainties in the structure of the progenitors and the failure to match the magnitude of the velocities of the incoming streams? Since the merging process is driven by a conversion of orbital angular momentum to internal angular momentum of the disks and dark halos (Barnes 1988, Hernquist 1992, 1993a), the initial mass distribution of the model galaxies will have a strong effect on the interaction dynamics and evolution of the tidal tails. Accordingly, tidal tails can potentially be used *in a statistical sense* as probes of the dark matter distribution around galaxies (Barnes 1988; Mihos, Dubinski, & Hernquist 1995). However, given the current uncertainties in the mass distribution of galaxy halos, the choice of dark matter structure for the galaxy models represents a significant uncertainty in the derivation of *specific* galaxy models.

  In spite of these uncertainties, the main results should remain qualitatively the same. The fact that we get a consistent match between the model and the observed kinematics, morphology, and timescales is an indication that the model provides a reasonable representation of the actual physical conditions. Therefore, while the derived quantities of tidal material are constrained only for a situation appropriate to our initial conditions, the delayed return of tidally lifted material seems firm. Likewise, the monotonic orbital structure of the tails, the occurrence of a fall-back radius, and the decreasing rate of return of tidally returning material should be a generic result of galaxy mergers.

  We would like to thank Lars Hernquist, Josh Barnes, and Kirk Borne for helpful discussions. JEH thanks Jacqueline van Gorkom, Francois Schweizer, David Helfand, R. Michael Rich, Puragra Guhathakurta and Dave Schiminovich for many useful comments on an earlier version of this paper. JEH also thanks Lars Hernquist and UCSC for their hospitality during the summer of 1993 to carry out this investigation. This work was supported in part by the San Diego Supercomputing Center, by the NSF under Grants AST 90–18526 and ASC 93–18185 to UCSC and Grants AST 89-17744 and AST 90-23254 to Columbia University, and by Grant HF–1059.01–94A from the Space Telescope Science Institute, which is operated by the Association of Universities for Research in Astronomy, Inc., under NASA contract NAS5–26555.

**FIGURE 6:**   The first 9 frames of Fig. 4, as viewed from above the orbital plane of the two galaxies $(X_o, Y_o)$. The particles from the NW disk are color coded according to their position at T=8. Magenta represents the innermost, most tightly bound disk material and accounts for 74% of the original NW disk. Blue is the material next farthest out and accounts for 12% of the disk material, followed by green (8%) and finally red (6%). Initial, luminous tails composed of 27% of the disk material are raised (blue, green, and red, T=16), but most of this falls back in quickly (blue, by T=32) . At the present time, only the red particles remain in the tails, and these continue to rain back upon the remnant.

**FIGURE 8:**   (a) The radial period versus pericentric distance for the tail particles, calculated according to their energies and angular momenta at T=72 (0.6 $h^{-1}$ Gyr). $R_{peri}$ is quantified in terms of $R_e$ by the red horizontal dashed lines. The particles are color coded according to their return times and radii. Magenta denotes particles with a pericentric distance of less than $4R_e$ and a period shorter than the time at the end of the simulation ($t_{end}$ =1.4 $h^{-1}$ Gyr). Blue and green particles are those which are predicted to return to within $4R_e$ and $5R_e$ respectively between 1.4 and 3 $h^{-1}$ Gyr. Red particles are expected to approach within $5R_e$ sometime between 3 $h^{-1}$ Gyr and 10 $h^{-1}$ Gyr of orbital periapse. Outer particles which will never fall to within $5R_e$ are drawn as black points. The large black dot is drawn at the location of the observed clump in the NW tail (HGvGS). The plot predicts that this clump will fall to within 14 $h^{-1}$ kpc in 3.1 $h^{-1}$ Gyr. (b) The sky view of the simulation at T=72, with the tail particles color coded as in (a). This figure is used to construct a "mask" used on the observational data to derived the quantities listed in Table 1.